\pdfoutput=1

\documentclass[11pt,a4paper]{article}
\usepackage{naacl2021}
\usepackage{times}
\usepackage{latexsym}

\usepackage[T1]{fontenc}

\usepackage[utf8]{inputenc}

\usepackage{amsmath}
\usepackage{graphicx}
\usepackage{subfig}
\usepackage{multirow}
\usepackage{booktabs}
\usepackage{fdsymbol}
\usepackage{url}
\usepackage{siunitx, makecell, caption}
\usepackage{balance}
\usepackage{makecell}
\usepackage[inline]{enumitem}
\usepackage{xcolor}
\usepackage{todonotes}
\presetkeys{todonotes}{inline}{}

\usepackage{microtype}

\newcommand\blfootnote[1]{%
  \begingroup
  \renewcommand\thefootnote{}\footnote{#1}%
  \addtocounter{footnote}{-1}%
  \endgroup
}

\title{Open-Domain Question Answering Goes Conversational\\via Question Rewriting}

\author{Raviteja Anantha\textsuperscript{$\medstar\vardiamondsuit$}, Svitlana Vakulenko\textsuperscript{$\medstar\spadesuit\varheartsuit$}, Zhucheng Tu\textsuperscript{$\vardiamondsuit$}, Shayne Longpre\textsuperscript{$\vardiamondsuit$},\\ \textbf{Stephen Pulman}\textsuperscript{$\vardiamondsuit$}, \textbf{Srinivas Chappidi}\textsuperscript{$\vardiamondsuit$} \\ 
\textsuperscript{$\vardiamondsuit$}{Apple Inc.}\\
\textsuperscript{$\spadesuit$}{University of Amsterdam, Amsterdam, the Netherlands}\\
}

\date{}

\begin{document}
\maketitle
\begin{abstract}
We introduce a new dataset for \textbf{Q}uestion \textbf{Re}writing in \textbf{C}onversational \textbf{C}ontext (QReCC), which contains 14K conversations with 80K question-answer pairs.
The task in QReCC is to find answers to conversational questions within a collection of 10M web pages (split into 54M passages).
Answers to questions in the same conversation may be distributed across several web pages.
QReCC provides annotations that allow us to train and evaluate individual subtasks of question rewriting, passage retrieval and reading comprehension required for the end-to-end conversational question answering (QA) task.
We report the effectiveness of a strong baseline approach that combines the state-of-the-art model for question rewriting, and competitive models for open-domain QA.
Our results set the first baseline for the QReCC dataset with F1 of 19.10, compared to the human upper bound of 75.45, indicating the difficulty of the setup and a large room for improvement.
\end{abstract}


\blfootnote{$\medstar$ Equal contribution.}
\blfootnote{$\varheartsuit$ Work done as an intern at Apple Inc.}

\section{Introduction}
\label{intro}

It is often not possible to address a complex information need with a single question.
Consequently, there is a clear need to extend open-domain question answering (QA) to a conversational setting.
This task is commonly referred to as conversational (interactive or sequential) QA~\cite{Webb:2006:1641579,DBLP:conf/emnlp/SaeidiBL0RSB018,DBLP:journals/tacl/ReddyCM19}.
Conversational QA requests an answer conditioned on both the question and the previous conversation turns as context.
Previously proposed large-scale benchmarks for conversational QA, such as QuAC and CoQA, limit the topic of conversation to the content of a single document.
In practice, however, the answers can be distributed across several documents that are relevant to the conversation, or the topic of the conversation may also drift.
To investigate this phenomena and develop approaches suitable for the complexities of this task, we introduce a new dataset for open-domain conversational QA, called QReCC.\footnote{\url{https://github.com/apple/ml-qrecc}}
The dataset consists of 13.6K conversations with an average of 6 turns per conversation.

\begin{figure}[t!]
  \includegraphics[width=\columnwidth]{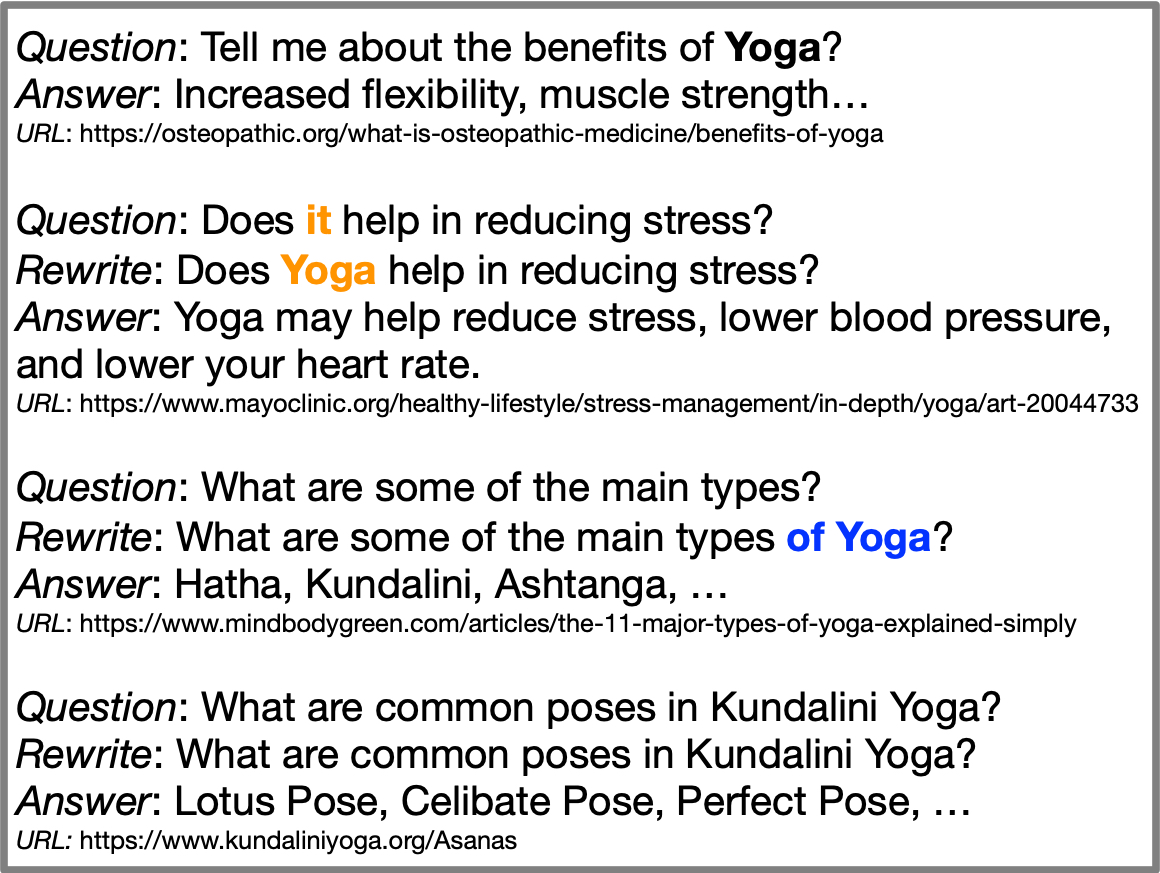}
  \caption{A snippet of a sample conversation from QReCC with question rewrites and answer provenance links. \textcolor{orange}{\textbf{Orange}} indicates coreference cases where the highlighted token should be replaced with its antecedent (in \textbf{bold}). \textcolor{blue}{\textbf{Blue}} indicates the tokens that should be generated to make the question unambiguous outside of the conversational context.}
  \label{fig:sampleconv}
\vspace*{-0.25in}
\end{figure}

A conversation in QReCC consists of a sequence of question-answer pairs.
The answers to questions were produced by human annotators, who looked up relevant information on the web using a search engine.
QReCC is therefore the first large-scale dataset for conversational QA that incorporates an information retrieval subtask.
QReCC is accompanied with scripts for building a collection of passages from the Common Crawl and the Wayback Machine for passage retrieval.

QReCC is inspired by the task of question rewriting (QR) that allows us to reduce the task of conversational QA to non-conversational QA by generating self-contained versions of contextually-dependent questions.
QR was recently shown crucial for porting retrieval QA architectures to a conversational setting~\cite{dalton2019trec}.
Follow-up questions in conversational QA often depend on the previous conversation turns due to ellipsis (missing content) and coreference (anaphora).
Every question-answer pair in QReCC is also annotated with a question rewrite. We evaluate the  quality of these rewrites as self-contained questions in terms of the ability of the rewritten question, when  used as input to the web search engine, to retrieve the correct answer.
A snippet of a sample QReCC conversation is given in Figure~\ref{fig:sampleconv}.

The dataset collection included two phases:
\begin{enumerate*}[label=(\arabic*)]
    \item dialogue collection, and
    \item document collection.
\end{enumerate*}
First, we set up an annotation task to collect dialogues with question-answer pairs along with question rewrites and answer provenance links.
Second, after all dialogues were collected we downloaded the web pages using the provenance links, and then extended this set with a random sample of other web pages from Common Crawl, preprocessed and split the pages into passages.

To produce the first baseline, we augment an open-domain QA model with a QR component that allows us to extend it to a conversational scenario.
We evaluate this approach on the QReCC dataset, reporting the end-to-end effectiveness as well as the effectiveness on the individual subtasks separately.

\paragraph{Our contributions.}
We collected the first large-scale dataset for end-to-end, open-domain conversational QA that contains question rewrites that incorporate conversational context.
We present a systematic comparison of existing automatic evaluation metrics on assessing the quality of question rewrites and show the metrics that best correlate with human judgement.
We show empirically that QR provides a unified and effective solution for resolving references --- both co-reference and ellipsis --- in multi-turn dialogue setting and positively impacts the conversational QA task.
We evaluate the dataset using a baseline that incorporates the state-of-the-art model in QR and competitive models for passage retrieval and answer extraction. 
This dataset provides a resource for the community to develop, evaluate, and advance methods for end-to-end, open-domain conversational QA.

\section{Related Work}
\label{relwork}
QReCC builds upon three publicly available datasets and further extends them to the open-domain conversational QA setting: Question Answering in Context (QuAC)~\cite{DBLP:conf/emnlp/ChoiHIYYCLZ18}, TREC Conversational Assistant Track (CAsT)~\cite{dalton2019trec} and Natural Questions (NQ)~\cite{47761}.
QReCC is the first large-scale dataset that supports the tasks of QR, passage retrieval, and reading comprehension (see Table~\ref{related_datasets} for the dataset comparison).


\begin{table*}
\centering
\caption{\label{related_datasets} The datasets that QReCC extends to open-domain conversational QA (QuAC, CAsT and NQ) and the datasets that are complementary to QReCC (CANARD and SaaC). RC - Reading Comprehension, PR - Passage Retrieval, QR - Question Rewriting.}
\begin{tabular}{lrrcc}
\hline \textbf{Dataset}                                                         & \textbf{\#Dialogues}               & \textbf{\#Questions}               & \textbf{Task}                    & \textbf{Provenance}        \\ \hline
QuAC~\cite{DBLP:conf/emnlp/ChoiHIYYCLZ18} & 13.6K                     & 98K                       & RC           & -          \\
NQ~\cite{47761}                          & 0                         & 307K                      & RC           & -          \\
CAsT~\cite{dalton2019trec}                & 80                        & 748                       & PR                      & -          \\
\hline
CANARD~\cite{DBLP:conf/emnlp/Elgohary19}  & 5.6K & 41K                     & QR                      & QuAC              \\
OR-QuAC~\cite{chen2020sigir}  & 5.6K & 41K                     & PR+RC                      & QuAC+CANARD              \\
SaaC~\cite{ren2020conversations} & 80                        & 748                       & QR+PR+RC & CAsT              \\
\hline
QReCC (our work)                                               & 13.7K & 81K & QR+PR+RC & QuAC+NQ+CAsT \\
\hline
\end{tabular}
\vspace*{-0.2in}
\end{table*}

\paragraph{Open-domain QA.}
Reading comprehension (RC) approaches were recently extended to incorporate a retrieval subtask~\cite{chen2017reading,yang2019end,lee2019latent}.
This task is also referred to as machine reading at scale~\cite{chen2017reading} or end-to-end QA~\cite{yang2019end}.
In this setup a reading comprehension component is preceded by a document retrieval component.
The answer spans are extracted from documents retrieved from a document collection, given as input.
The standard approach to end-to-end open-domain QA is
\begin{enumerate*}[label=(\arabic*)]
    \item use an efficient filtering approach to reduce the number of candidate passages to the top-$k$ of the most relevant ones (usually BM25 based on the bag-of-words representation); and then
    \item re-rank the subset of the top-$k$ relevant passages using a more fine-grained approach, such as BERT based on vector representations~\cite{yang2019end}.
\end{enumerate*}

\paragraph{Conversational QA.}
Independently from end-to-end QA, the RC task was extended to a conversational setting, in which answer extraction is conditioned not only on the question but also on the previous conversation turns~\cite{DBLP:conf/emnlp/ChoiHIYYCLZ18,DBLP:journals/tacl/ReddyCM19}.
The first attempt at extending the task of information retrieval (IR) to a conversational setting was the recent TREC CAsT 2019 task~\cite{dalton2019trec}.
The challenge was to rank passages from a passage collection by their relevance to an input question in the context of a conversation history.
The size of the collection in CAsT 2019 was 38.4M passages, requiring efficient IR approaches.
As efficient retrieval approaches operate on bag-of-words representations they need a different way to handle conversational context since they can not be trained end-to-end using a latent representation of the conversational context.
A solution to this computational bottleneck was a QR model that learns to sample tokens from the conversational context as a pre-processing step before QA.

\paragraph{Question Rewriting.}
CANARD~\cite{DBLP:conf/emnlp/Elgohary19} provides rewrites for the conversational questions from the QuAC dataset.
QR effectively modifies all follow-up questions such that they can be correctly interpreted outside of the conversational context as well.
This extension to the conversational QA task proved especially useful while allowing retrieval models to incorporate conversational context~\cite{voskarides-2020-query,vakulenko2020question,lin2020query}.

More recently, \citeauthor{chen2020sigir} introduced OR-QuAC dataset that was automatically constructed from QuAC and CANARD datasets. OR-QuAC uses the same rewrites and answers as the ones provided in QuAC and CANARD.
In contrast to OR-QuAC, the answers in QReCC are not tied to a single Wikipedia page. The answers can be distributed across several web pages. QReCC's passage collection is also larger and more diverse: 11M passages from Wikipedia in OR-QuAC vs. 54M passages from CommonCrawl in QReCC. The answers in OR-QuAC are single spans, whereas QReCC answers were produced by human annotators instructed to imitate natural conversational answers and may include several spans from different parts of the same web page.

TREC CAsT 2019 paved the way to conversational QA for retrieval but had several important limitations:
\begin{enumerate*}[label=(\arabic*)]
    \item no training data and
    \item no answer spans.
\end{enumerate*}
First, the size of the CAsT dataset is limited to 80 dialogues, which is nowhere enough for training a machine-learning model.
This was also the reason why CANARD played such an important role for the development of retrieval-based approaches even though it was collected as a RC dataset.
Second, the task in TREC CAsT 2019 was conversational passage retrieval not extractive QA since the expected output was ranked passages and not a text span.
We designed QReCC to overcome both of these limitations.

The size of the QReCC dataset is comparable with other large-scale conversational QA datasets (see Table~\ref{related_datasets}).
The most relevant to our work is the concurrent work by \citeauthor{ren2020conversations}, who extended the TREC CAsT dataset with crowd-sourced answer spans.
Since the size of this dataset is inadequate for training a machine-learning model and can be used only for evaluation, the authors train their models on the MS MARCO dataset instead, which is a non-conversational QA dataset~\cite{bajaj2016msmarco}.
Their evaluation results show how the performance degrades due to the lack of conversational training data.
TREC CAsT will continue in the future and the QReCC dataset provides a valuable benchmark helping to train and evaluate novel conversational QA approaches.

\section{Dialogue Collection}
To simplify the data collection task we decided to use questions from pre-existing QA datasets as seeds for dialogues in QReCC.
We used questions from QuAC, CAsT and NQ.
While QuAC and CAsT datasets contain question sequences, NQ is not a conversational dataset but contains stand-alone questions from web search.
We use the NQ dataset to increase and diversify the number of samples beyond QuAC and CAsT by generating more rewrites for cases beyond coreference resolution.
The majority of the follow-up questions in QuAC require coreference resolution for QR.
Therefore, we explicitly instructed the annotators to use NQ as a start of a conversation and then come up with relevant follow-up questions, which would require generation of missing content, i.e., ellipsis, instead of coreference resolution for QR.

\begin{table}[t]
\caption{Summary statistics for the QReCC dataset.}
  \scalebox{0.9}{
\begin{tabular}{lllll}
  \toprule
QReCC & Train & Dev. & Test & All \\
 \toprule
\# questions (Qs)& 50.8K  & 12.7K & 16.4K & 80.0K \\
\# dialogues & 8.7K & 2.2K  & 2.8K  & 13.6K \\
max Qs/dialogue & 12 & 12 & 12 & 12 \\
avg Qs/dialogue & 6 & 6 & 6 & 6\\
min Qs/dialogue & 5 & 5 & 5 & 5 \\
\midrule
\% replacement & 53 & 52 & 53 & 52 \\
\% insertion & 35 & 36 & 37 & 38 \\
\% copy & 11 & 11 & 9 & 9 \\
\% removal & 1 & 1 & 1 & 1 \\
\bottomrule
\end{tabular}
}
\label{tab:dataset_stats}
\end{table}

The task for the annotators was also to answer questions using a web search engine.
Question rewrites were used as input to a search engine.
This setup helps to obtain feedback on the quality of QR with respect to the effectiveness of answer retrieval (see Section~\ref{sec:metric} for more details on using search results for the evaluating QR).
Finally, the question-answer pair is annotated with the link to the web page that was used to produce the answer.

Thereby, every dialogue was produced by the same annotator including the questions, answers and rewrites. This design decision is called \textit{self-dialog technique} that was shown to help improve quality of the data by avoiding some of the challenges observed in simulated dialogues produced by pairs of annotators~\cite{byrne2019taskmaster}.

A team of 30 professional annotators with a project lead were employed to perform the task.
The annotation task was described in the guidelines (see Appendix~\ref{appendix:guidelines} for more details).
To ensure the quality of the annotations we followed a post-hoc evaluation procedure, in which 5 reviewers go through the dataset and update incorrect examples they identify with consensus.

\section{Dialogue Analysis}
\label{sec:analysis}
QReCC contains 13,598 dialogues with 79,952 questions in total.
9.3K dialogues are based on the questions from QuAC; 80 are from TREC CAsT; and 4.4K are from NQ.
9\% of questions in QReCC do not have answers.
We still retained the question rewrites even if no answer was found on the web.
112 questions were annotated with links to web pages without answer texts, e.g. ``May I have a link to road signs in Singapore?''

We prepared three standard dataset splits and ensured that they are balanced in terms of the standard dialogue statistics and the types of QR (see Table~\ref{tab:dataset_stats}).
We distinguish four types of QR.
They differ with respect to the intervention required to resolve contextual dependencies in dialogue.
These types can be automatically identified by measuring the difference between an original question $Q$ and a question rewrite $R$ that are represented as sets using the bag-of-words:
\begin{itemize}
\item \textit{Insertion} -- new tokens are added to the original question to produce the rewrite (e.g., ``What are some of the main types'' $\rightarrow$ ``What are some of the main types \textit{of Yoga}?''):\\
$Q \setminus R = \emptyset \land R \setminus Q \ne \emptyset$
\item \textit{Removal} -- some tokens are removed from the question to produce the rewrite (e.g., ``Can you tell me about the C++ language mentioned'' $\rightarrow$ ``Can you tell me about the C++ language''):\\
$Q \setminus R \ne \emptyset \land R \setminus Q = \emptyset$
\item \textit{Replacement} -- some tokens are added and some are removed to produce the rewrite (e.g., ``Does it help in reducing stress'' $\rightarrow$ ``Does \textit{Yoga} help in reducing stress''):\\
$Q \setminus R \ne \emptyset \land R \setminus Q \ne \emptyset$
\item \textit{Copy} -- no modification is needed, i.e., the original question is already contextually independent (e.g., ``What are common poses in Kundalini Yoga?''):\\
$ Q \setminus R = \emptyset \land R \setminus Q = \emptyset$, i.e., $Q = R$
\end{itemize}

\begin{figure}[t]
  \centering
  \includegraphics[width=\linewidth]{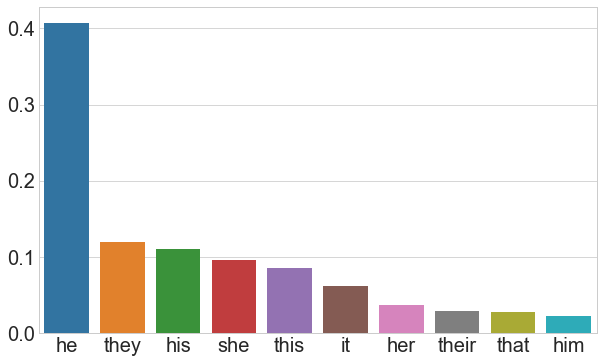}
  \caption{The 10 most frequently replaced tokens in QReCC.}
  \label{frequent_tokens}
 \vspace*{-0.2in}
\end{figure}

The majority of questions in QReCC (52\%) require \textit{Replacement}. 
Figure~\ref{frequent_tokens} shows the tokens that are most frequently replaced in QR.
All of them are pronouns that require anaphora resolution.
By specifically targeting more rare types of question rewriting in our data collection task we managed to increase the proportion of the \textit{Insertion} cases in our dataset.
This allows us to train and evaluate the ability of the model to reconstruct missing context, which cannot be achieved using traditional co-reference resolution approaches.

\section{Document Collection}

\begin{table*}[t]
  \centering
  \caption{Comparison of different evaluation metrics in terms of Pearson correlation with the human judgment of the question rewriting quality.}
  \label{tab:metrics}
\begin{tabular}{llc|lc}
   \toprule
\multicolumn{2}{l}{Metrics} & Pearson & Metrics & Pearson \\
    \midrule
\multicolumn{2}{l}{Exact Match} & 0.56 & ROUGE-1 P & 0.51 \\
Embeddings & \textbf{USE} & \textbf{0.67} & \textbf{ROUGE-1 R} & \textbf{0.63} \\
 & InferSent & 0.48 & ROUGE-1 F & 0.61 \\
\multirow{2}{*}{\shortstack[l]{Search\\ Results}} & R@1 & 0.66 & ROUGE-2 P & 0.54 \\
 & R@2 & 0.72 & ROUGE-2 R & 0.57 \\
 & R@3 & 0.73 & ROUGE-2 F & 0.57 \\
 & R@4 & 0.74 & ROUGE-L P & 0.50 \\
 & R@5 & 0.77 & ROUGE-L R & 0.61 \\
 & \textbf{R@10} & \textbf{0.80} & ROUGE-L F & 0.58 \\
 & AR & 0.79 & METEOR & 0.59 \\
 & NDCG & 0.74 & BLEU & 0.58 \\
  \bottomrule
\end{tabular}
\end{table*}

We download the web pages using the answer provenance links provided by the annotators from the Internet Archive Wayback Machine.\footnote{We use the version of a web page, which is the closest to the end date of the dialogue collection (November 24, 2019).}
Then, we complement the relevant pages with randomly sampled web pages that constitute 1\% of the Common Crawl dataset identified as English pages. 
The final collection consists of approximately 14K pages from the Wayback Machine and 9.9M random web pages from the Common Crawl dataset.
The scripts for reproducing the passage collection are on GitHub.
See Appendix~\ref{app:doc_collection} for more details.

After downloading the pages we extract the textual content from the HTML and split texts into passages of least 220 tokens.
After segmentation, we have a total of 54M passages which we index using Anserini \cite{yang2017anserini}.

We search the passage collection using the human annotated answers to augment the dataset with alternative sources of correct answers.
For each document returned, we identify the span in the document that has the highest token overlap (F1) with the human answer.
We consider all documents with F1 $\geq$ 0.8 as relevant.
Verifying adequacy of this simple heuristic by human annotators is left for future work.

\section{Question Rewriting Metrics Validation}
\label{sec:metric}

\begin{table*}[t]
\caption{\label{tab:qr_results} Evaluation results of QR models (mean with 95\% confidence intervals). *Human QR metrics are computed across 5 different random samples of 1000 question rewrites from the intersection of QReCC and CANARD conversations.}
  \centering
 \scalebox{0.92}{
\begin{tabular}{l|ccc}
   \toprule
Model/Metrics & ROUGE-1 R & USE & R@10 \\
    \midrule
AllenAI Coref~\cite{DBLP:conf/naacl/LeeHZ18} &  67.1\% $\pm$ 10E-4\%  &  82.3\% $\pm$ 10E-3\%  &  56.1\% $\pm$ 10E-4\% \\
Generator~\cite{radford2019language} &  73.4\% $\pm$ 0.6\% &  86.2\% $\pm$ 0.9\%  &  69.1\% $\pm$ 0.2\% \\
Generator + Multiple-choice~\cite{DBLP:journals/corr/abs-1901-08149} &  74.1\% $\pm$ 0.5\%  &  86.3\% $\pm$ 0.4\% & 70.2\% $\pm$ 0.1\% \\
PointerGenerator~\cite{DBLP:conf/emnlp/Elgohary19} &  80.2\% $\pm$ 0.8\% &  89.1\% $\pm$ 1.1\% &  75.3\% $\pm$ 0.3\% \\ 
GECOR~\cite{DBLP:conf/emnlp/gecor} &  84.1\% $\pm$ 0.3\% &  91.8\% $\pm$ 0.2\%  &  78.1\% $\pm$ 0.2\%  \\
CopyTransformer~\cite{DBLP:conf/emnlp/GehrmannDR18} &  86.1\% $\pm$ 0.5\%  &  92.8\% $\pm$ 0.3\%  &  79.4\% $\pm$ 0.3\%  \\
Transformer++ &  \textbf{89.5\% $\pm$ 0.4\%}  &  \textbf{95.2\% $\pm$ 0.2\%}  &  \textbf{83.2\% $\pm$ 0.3\%}  \\
     \midrule
Human* & 94.6\% $\pm$ 0.2\% & 97.3\% $\pm$ 0.1\% & 87.2\% $\pm$ 0.1\% \\
  \bottomrule
\end{tabular}}
\vspace*{-0.2in}
\end{table*}

BLEU has typically been used in previous work for measuring the quality of QR~\cite{DBLP:conf/emnlp/Elgohary19,lin2020query}.
We conduct a systematic evaluation and compare BLEU with alternative metrics, previously applied in summarization and translation, to ensure the most reliable metrics we can obtain for the model selection.
Our evaluation shows that BLEU does not compare favourably with other metrics in evaluating the quality of QR.

\paragraph{Task.}
We took a random sample of 10K questions and used a seq-to-seq model~\cite{DBLP:journals/acl/Nallapati16} trained with questions and conversation context from the QReCC dataset to generate question rewrites.
These generated rewrites were compared to the ground truth rewrites produced by human annotators.
Different annotators graded each model-generated rewrite with a binary label: 0 (incorrect rewrite) or 1 (correct rewrite).
For a question rewrite to be correct it does not have to exactly match the ground truth rewrite, but it should correctly capture the conversational context and be a self-contained question.
For example, the model-generated rewrite ``What are the global warming dangers?'' is a correct rewrite with the ground truth rewrite being ``What are the dangers of global warming?''. 
In addition, we also assess the variance of the human assessments. The Pearson correlation between any two annotators on average is 0.94. We observed the mean and the variance to be 0.083 and 0.076 respectively. Performing a two-tail statistical significance test shows the P-value to be 0.0201.

We use several automated metrics to compare the rewrites with the ground truth and compute their Pearson correlation with the human judgements (see Table~\ref{tab:metrics} for results).\\

\textbf{Exact Match} is a binary variable that indicates the token set overlap applied after the standard preprocessing: lower-casing, stemming, punctuation and stopword removal.\\

\textbf{ROUGE}~\cite{lin2004rouge} reflects similarity between two texts in terms of n-gram overlap (R-1 for unigrams; R-2 for bigrams and R-L for the longest common n-gram). We report the mean for precision (P), recall (R) and F-measure (F).\\

\textbf{METEOR}~\cite{DBLP:conf/wmt/DenkowskiL14} is a machine translation metric based on exact, stem, synonym, and paraphrase matches between words and phrases.\\

\textbf{BLEU}~\cite{DBLP:conf/emnlp/Papineni02} is a text similarity metric that uses a modified form of precision and n-grams from candidate and reference texts.\\

\textbf{Embeddings} group several unsupervised approaches that produce a sentence-level vector representation: Universal Sentence Encoder~\cite{DBLP:conf/emnlp/CerYKHLJCGYTSK18} and InferSent~\cite{conneau-EtAl:2017:EMNLP2017}.\\

\textbf{Search Results} -- we use both question rewrites in Google Search and compare the overlap between the produced page ranks in terms of the standard IR metrics: Recall@$k$ for the top-$k$ links, Average Recall (AR) and Normalized Discounted Cumulative Gain (NDCG).\\






The best performing metric in our experiments (i.e., closest to the human judgement) is the set overlap of the web search results (\textbf{R@10}).
The best metrics independent of QA are Universal Sentence Embedding (\textbf{USE}) and unigram recall (\textbf{ROUGE-1 R}).
We provide more details of the metrics performance illustrated with examples and the discussion in Appendix~\ref{appendix:metrics}.
We use the set of all three best evaluation metrics to select the optimal QR model for our baseline approach.

\section{Baseline Approach}

\begin{figure*}[t]
 \centering
  \includegraphics[scale=0.063]{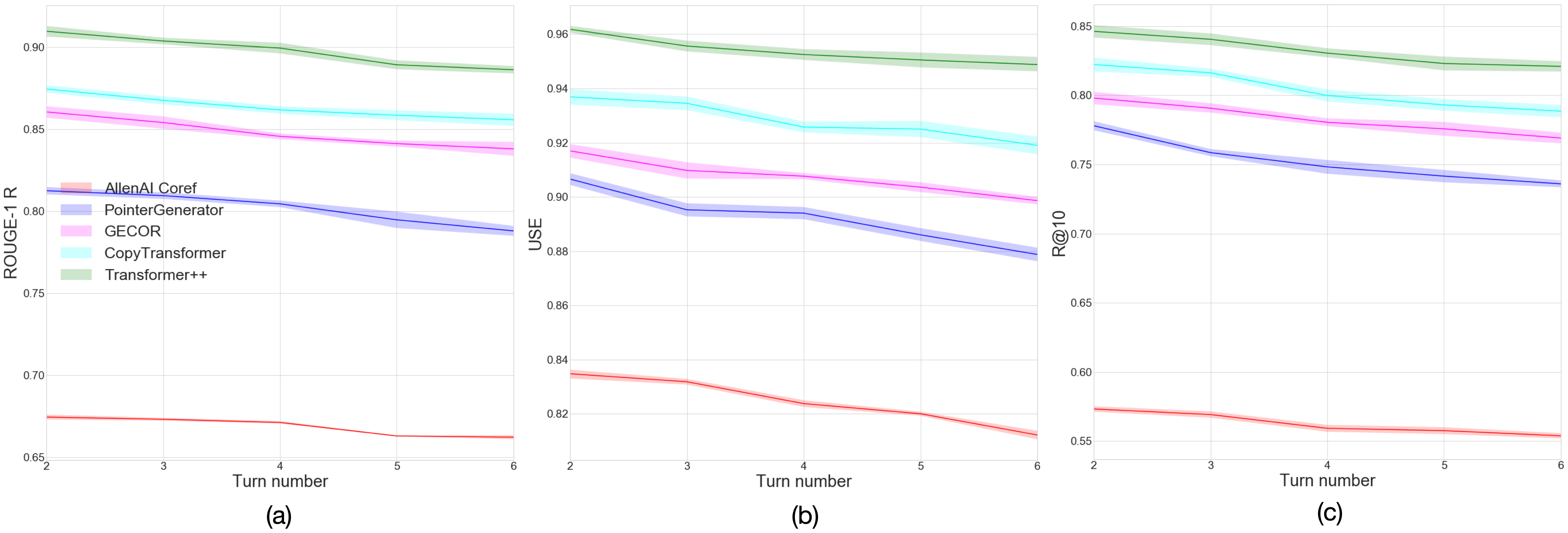}
  \caption{Rouge-1R, USE and R@10 metrics of baseline co-reference model, top-3 encoder-decoder models, and Transformer++ model based on dialogue turn number}
  \label{fig:turns}
\vspace*{-0.2in}
\end{figure*}

We extend BERTserini~\cite{yang2019end}, an efficient approach to open-domain QA, with a QR model to incorporate conversational context.
This approach consists of three stages: (1) QR, (2) PR and (3) RC.
First, a model is trained to generate a stand-alone question given a follow-up question and the preceding question-answer pairs.
In the second stage, PR, the top-$k$ relevant passages are retrieved from the index using BM25 using the rewritten question.
Finally, in RC, a model is trained to extract an answer span from a passage or predict if the passage is irrelevant.
The scores obtained from PR and RC are then combined as a weighted sum to produce the final score.
The span with the highest score is chosen as the final answer.

\subsection{Question Rewriting}
\label{label:qr}

We evaluate a co-reference model and several generative models on the QR subtask using the question rewrites in QReCC and the set of QR metrics selected in Section~\ref{sec:metric}.
The best performing model is then used in a combination with BERTserini to set the baseline results for the end-to-end QA task.
All our Transformer-based models were initialized with the pretrained weights of GPT-2 (English medium-size)~\cite{radford2019language} and further fine-tuned on question rewrites from the QReCC training set (see Appendix~\ref{app:QR}).\\

\textbf{AllenAI Coref} is the state-of-the-art model for coreference resolution task~\cite{DBLP:conf/naacl/LeeHZ18}. We adapt it for QR with a heuristic that substitutes all coreference mentions with the corresponding antecedents from the cluster.\\

\textbf{PointerGenerator} uses a bi-LSTM encoder and a pointer-generator decoder, which allows to copy and generate tokens~\cite{DBLP:conf/emnlp/Elgohary19}.\\

\textbf{GECOR} uses two bi-GRU encoders, one for user utterance and other for dialogue context, and a pointer-generator decoder previously proposed for task-oriented dialogues~\cite{DBLP:conf/emnlp/gecor}. \\

\textbf{Generator} is a Transformer decoder model with a language modeling head (linear layer in the size of the vocabulary)~\cite{radford2019language}.\\

\textbf{Generator + Multiple-choice} model has a second head for the auxiliary classification task that distinguishes between the correct rewrite and several noisy rewrites as negative samples (inspired by TransferTransfo~\cite{DBLP:journals/corr/abs-1901-08149}).\\

\textbf{CopyTransformer} uses one of the attention heads of the Transformer as a pointer to copy tokens from the input sequence directly~\cite{DBLP:conf/emnlp/GehrmannDR18}.\\

\textbf{Transformer++} model has two language modeling heads that produce separate vocabulary distributions, which are then combined via a parameterized weighted sum (the coefficients are produced by combining the output of the first attention head and the input embeddings).

\subsection{BERTserini}
We implemented BERTserini following \citet{yang2019end}
We use the standard BM25 ranking for passage retrieval with $k_1=0.82$, $b=0.68$, which was previously found to work well for passage retrieval on MS MARCO.
We then retrieve the top-100 relevant passages per question.
Afterwards, we use BERT-Large fine-tuned for the task of reading comprehension.
This model takes a question and each of the relevant passages as input and produces the answer span~\cite{Wolf2019HuggingFacesTS}.
BERT-Large produces a score ($S_{\text{BERT}}$), which is combined with the retrieval score for each of the passages ($S_{\text{Anserini}}$) through simple linear interpolation:
\begin{align*}
    S = (1 - \mu) \cdot S_{\text{Anserini}} + \mu \cdot S_{\text{BERT}}
\end{align*}

We pick the span with the highest score $S$ as the answer.
The parameter $\mu \in [0, 1]$ was tuned using a 10\% random subset of the QReCC training set withheld from the BERT-Large training (we found $\mu = 0.7$ to work best).

BERT-Large was trained on human rewrites from the QReCC training set, and evaluated on the test set using either the original questions, human rewrites or the rewrites produced by Transformer++.
The model is trained to either predict an answer span or predict that the passage does not contain an answer. 
``No answer'' for the question is predicted only when neither of the relevant passages predicts an answer span.
The model was trained on 480K paragraphs that contain the correct answers and 5K of other paragraphs as negative samples (see Appendix~\ref{app:QA} for more details).

\section{Baseline Results}

We use the results of QR to select the best model and then use it for the end-to-end QA task.
Question rewrites are used as input for both passage retrieval and reading comprehension tasks.
The effectiveness of the QR component is compared with the end-to-end model conditioned on the conversational context.

\subsection{Question Rewriting Effectiveness}

We analyze the effectiveness of our QR models by doing a 5-fold cross validation and obtaining the best performing metrics.
Figure~\ref{fig:turns} contains 3 plots showing ROUGE 1-R, USE and R@10 across 5 turns. We start with the second turn because the first turn always is a self-contained query.
The metrics across turns also stay stable with the same result for all the models. The Transformer++ model is stable with little variance in terms of its maximum and minimum metric values across all the best performing metrics.

Our evaluation results are summarized in Table~\ref{tab:qr_results}.
All generative models outperform the state-of-the-art coreference resolution model (AllenAI Coref). We noticed that PointerGenerator which employs a bi-LSTM encoder with a copy and generate mechanism outperforms Generator using Transformer alone. 
We could not find evidence that pretraining with an auxiliary regression task can improve the QR model effectiveness (Generator + Multiple-choice).
Use of two separate bi-GRU encoders for the query and conversation context further improved the QR effectiveness (GECOR).  
Modeling both copying and generating the tokens from the input sequence employing the Transformer helped improve the effectiveness of the QR model (CopyTransformer) compared to other existing generative models.
Finally, obtaining the final distribution by computing token probabilities and weighting question and context vocabulary distributions with those probabilities helped improve over the best performing generative model (Transformer++).

\subsection{Question Answering Effectiveness}

\begin{table}[t]
\caption{Mean reciprocal rank, recall@10, and recall@100 for passage retrieval on test set questions.}
\label{tab:retrieval_qa_results}
\centering
	\resizebox{\columnwidth}{!}{
\begin{tabular}{lrrr}
  \toprule
Rewrite Type & MRR & R@10 & R@100 \\
 \toprule
Original  & 0.0343 & 6.12 & 11.71 \\
Transformer++ & 0.1586 & 26.52 & 41.51 \\
Human & 0.1994 & 32.78 & 49.36 \\
\bottomrule
\end{tabular}}
\vspace*{-0.1in}
\end{table}
\begin{table}[t]
\caption{Mean F1 and Exact Match scores (\%) on passages for extractive QA. 
``Known Context" assumes perfect retrieval.
The ``Extractive Upper Bound" assumes perfect single document span extraction.
}
\label{tab:extractive_qa_results}
\centering
    \resizebox{\columnwidth}{!}{
\begin{tabular}{llrr}
  \toprule
Setting & Rewrite Type & F1 & EM \\
 \toprule
End-to-End & Original  & 9.07 & 0.32 \\
& Transformer++ & 19.10 & 1.01 \\
& Human & 21.82 & 1.23 \\
\midrule
Known Context & Original  & 17.24 & 1.90 \\
& Transformer++ & 32.34 & 4.04 \\
& Human & 36.42 & 4.70 \\
\midrule
\multicolumn{2}{l}{Extractive Upper Bound} & 75.45 & 25.07 \\
\bottomrule
\end{tabular}}
\vspace*{-0.2in}
\end{table}


Table~\ref{tab:retrieval_qa_results} shows the mean reciprocal rank (MRR), R@10, and R@100 of using the original, Transformer++, and human rewritten questions.
R@$k$ is averaged across all questions.
For a question, if R@$k$ is 1.0, it means that there is a passage in the top-$k$ at any rank such that the passage is relevant; and 0.0 otherwise.
Table~\ref{tab:extractive_qa_results} shows the standard F1 and Exact Match metrics for extractive QA for each type of input question.
In the ``End-to-End'' setting, the retrieval score was combined with the BERT reader score to determine the final span.
In the ``Known Context'' setting, we use the relevant passage from the web page indicated by the human annotator, i.e., without passage retrieval.
In the ``Extractive Upper Bound'' setting, we use a heuristic to find the answer span with the highest F1 score among the top-100 retrieved passages with human rewrite.
This setup indicates the best the reader can do given the retrieval results.

The upper bound on the answer span extraction (F1 $=$ 75.45) highlights the need for more sophisticated QA techniques than the standard reading comprehension approaches can offer now. 
Some answer texts in QReCC were paraphrased or summarised using multiple passages from the same web page.
Abstractive approaches to answer generation are necessary to close this gap.

Even using single document span extraction techniques, there is a large room for improvement. 
Comparing ``Known Context'' to ``End-to-End'' we see losses introduced by the retrieval step, and comparing the ``Extractive Upper Bound'' to ``Known Context'' we see the sizeable margin of improvement available even for extractive models.
This shows that even with competitive baselines the QA tasks are all far from solved.

In both Table~\ref{tab:retrieval_qa_results} and ~\ref{tab:extractive_qa_results} we see that human rewritten questions more than double the effectiveness of using original questions.
In the absence of human rewritten questions, using Transfomer++ elevates the effectiveness of the QA tasks, getting it much closer to that proffered by human-level QR.
\section{Conclusion}
We introduced the QReCC dataset for open-domain conversational QA.
QReCC is the first dataset to cover all the subtasks relevant for conversational QA, which include question rewriting, passage retrieval and reading comprehension.
We also set the first end-to-end baseline results for QReCC by evaluating an open-domain QA model in combination with a QR model.
We presented a systematic comparison of existing automatic evaluation metrics on assessing the quality of question rewrites and show the metrics that best proxy human judgement. 
Our empirical evaluation shows that QR provides an effective solution for resolving both ellipsis and co-reference that allows to use existing non-conversational QA models in a conversational dialogue setting.
Our end-to-end baselines achieve an F1 score of 19.10, well beneath the 75.45 extractive upper bound, suggesting not only room for improvement in extractive conversational QA, but that more sophisticated abstractive techniques are required to successfully solve QReCC.


\balance
\bibliography{anthology,naacl2021}
\bibliographystyle{acl_natbib}

\clearpage
\appendix

\section{Reproducibility}

\subsection{Training Transformer++ for Question Rewriting}
\label{app:QR}
Details about training setup of Transformer++ for question rewriting task is provided in Table~\ref{tab:transformer-hyperparams}. The Transformer head is initialized with the pretrained weights of GPT-2 (medium) and further fine-tuned on the QReCC train set.
We use PyTorch implementation from HuggingFace.\footnote{\url{https://github.com/huggingface/transformers}}
Transformer++ is trained using model parallelism on $5$ Tesla V100 GPUs with hyperparameter search trial.

\begin{table*}
\centering
\caption{\label{tab:transformer-hyperparams}
Hyperparameter selection and tuning ranges for \textsc{Transformer++} used for question rewriting.
}
\begin{tabular}{ll}
\toprule
\textsc{Model Parameters} & \textsc{Value/Range} \\
\midrule
\textbf{Fixed Parameters} & {} \\
\midrule
Batch Size & 16 \\
Optimizer & Adam \\
Vocabulary Size & 150,263 \\
Transformer Head & GPT-2 (medium) \\
Learning Rate Schedule & Exponential Decay \\
Output Attention & True \\
Max Input Sequence Length & 1024 \\
Max Output Sequence Length & 30 \\
Num Hyperparameter Search Trials & 500 \\
\midrule
\textbf{Tuned Parameters} & {} \\
\midrule
Num Epochs & [$50$, $100$] \\
Initializer Range & [$0.01$, $0.1$] \\
Dropout & [$0.05$, $0.2$] \\
Attention Dropout & [$0.05$, $0.1$] \\
Residual Dropout & [$0.05$, $0.1$] \\
Learning Rate & [$1e-3$, $1e-1$] \\
Decay Steps & [$6000$, $10000$] \\
Decay Rate & [$0.7$ , $0.9$] \\
Activation Functions & [ReLU, Leaky ReLU, GELU] \\
\midrule
\textbf{General} & {} \\
\midrule
Model Size (\# params) & $350$M \\
Avg. Train Time (per epoch) & $12$ hours \\
\bottomrule
\end{tabular}
\end{table*}

\subsection{Building Document Collection}
\label{app:doc_collection}
Here we provide further details for building the document collection.
If the web page of the provenance link containing the answer was not archived by the Wayback Machine yet, we trigger the archiving through the Wayback Machine API whenever possible.
Overall, 2\% of the annotated web pages could not be archived by the Wayback Machine due to the restricted access (such as the Quora website).

For the Common Crawl data, we take the index files from November 2019 and filter URLs to only those that are retrieved with HTTP status code 200 and those that are identified as English.
We extract the pages from the Common Crawl WET files that correspond to these filtered URLs, and sample the first link out of every 100 links in each filtered WET file.

Overall, we find that 97.8\% of unique web pages found by human annotators to contain answer and has an associated archived copy on the Wayback Machine.
The final collection consists of both these pages from the Wayback Machine and random web pages from the Common Crawl.

After downloading the pages we extract all text from the page using the Beautiful Soup library.\footnote{\url{https://www.crummy.com/software/BeautifulSoup/bs4/doc/}}
We iterate through the web page by newlines, and accumulate the tokens for every line.
Whenever the number of tokens reaches 220 or more, we emit a paragraph, and reset the token counter to 0.
Note the last paragraph on the page may have fewer than 220 tokens.
After segmentation, we have a total of 54,241,550 passages which we index using Pyserini 0.10.0.1.
Hence we treat each passage as a single document.

\subsection{Training BERT-L for Reading Comprehension}
\label{app:QA}
Below we provide details about the training setup for BERT-L used of the reading comprehension task in our experiments, which is similar to the extractive reader setup in \citet{longpre2019exploration} but using BERT-L. 
We train on the full data of the QReCC training set, using Human rewritten questions. 
Our implementation of the BERT question answering modules follows that of the standard PyTorch \citep{NEURIPS2019_9015} implementations from HuggingFace, and are trained on $4$ NVIDIA Tesla V100 GPUs.
The model is trained to predict an answer span or abstain if the passage has ``No Answer''. 
For every query we obtain up to 25 paragraphs from the document that contains the gold answer as identified by a human grader. 
The paragraph with the answer is always used for training, and a portion of the other paragraphs are used in training as No Answer or ``negative" examples.
Using the development set we tune several hyperparameters, most importantly the percentage of negative examples to retain for training (``Pct Neg. Ratio").
Fixed parameters and tuning details are shown in Table~\ref{tab:eqa-hyperparams}. 

\begin{table*}
\centering
\caption{\label{tab:eqa-hyperparams}
Hyperparameter selection and tuning ranges for \textsc{Bert-L} used for reading comprehension.
}
\begin{tabular}{ll}
\toprule
\textsc{Model Parameters} & \textsc{Value/Range} \\
\midrule
\textbf{Fixed Parameters} & {} \\
\midrule
Batch Size & 32 \\
Optimizer & Adam \\
Learning Rate Schedule & Exponential Decay \\
Num Epochs & 2 \\
Max Input Sequence Length & 512 \\
Max Span Length & 30 \\
Num Hyperparameter Search Trials & 32 \\
\midrule
\textbf{Tuned Parameters} & {} \\
\midrule
Learning Rate & [$1e-5$, $5e-5$] \\
Pct Neg. Ratio & [$0.01$, $0.5$] \\
\midrule
\textbf{General} & {} \\
\midrule
Model Size (\# params) & $330$M \\
Avg. Train Time (per epoch) & $8$ hours \\
\bottomrule
\end{tabular}
\end{table*}

\section{Annotation Guidelines}
\label{appendix:guidelines}

\paragraph{Instructions for question rewriting:}
\begin{itemize}
\item Rewritten questions should be as close to the original as possible.
\item Questions should not contain any references to the previous context of the conversation.
\item Avoid using any pronouns in question rewrites.
\end{itemize}

\paragraph{Instructions for answering questions:}
\begin{itemize}
\item Put the rewritten question (original question if it is already self-contained) in a web search engine to produce the correct answer. 
\item Produce an answer, which should be short and brief with minimum information required to answer the question.
\item The answers should be grammatically correct, do not contain special symbols or any additional mark-up.
\item Produce an answer that would be most natural for a human conversation.
\item Answers can contain up to a maximum of 30 words.
\item When the answer is not a text, provide the source URL only (e.g., a geo-location on a map or a music video link).
\end{itemize}

\section{Pitfalls of the Query Rewriting Metrics}
\label{appendix:metrics}

Our evaluation results show that the text similarity metrics, such as ROUGE and USE, often fall short to reflect semantic similarity in case of lexical paraphrases. Retrieval-based metrics, such as Recall@10, are able to demonstrate better correlation with human judgement.
However, retrieval-based metrics are more expensive to compute since it requires an API call for every query.
Also, they rely on the underlying collection as well as the ability of the search engine to handle paraphrases. Our experiments show that text similarity metrics, however flawed, are still able to provide a good proxy for quickly assessing QR performance and are suitable for comparing models in the development phase during parameter tuning.
Retrieval-based metrics are useful to better approximate human judgement but can be computed for the best models only that were pre-selected using text similarity metrics.\\

\textbf{ROUGE-1 R} metrics provides a very rough estimate of the model performance by counting the number of words missing from the generated question rewrite in comparison with the ground truth rewrite and does not have any mechanism to distinguish which words are more crucial than others.
As a result, a question missing only a single letter will receive the same score as a question missing one of its most informative words.
For example, ROUGE(``When is Robert Downey \textit{Jr} birthday'', ``When is Robert Downey \textit{Jrs} birthday'') = ROUGE(``When did Gabriel Garcia die'', ``When did Gabriel Garcia \textit{Marquez} die'') = 0.75.\\

\textbf{USE} is more sensitive to such variations and can better pick up on the character-level similarities: compare to USE(``When is Robert Downey \textit{Jr} birthday'', ``When is Robert Downey \textit{Jrs} birthday'')=0.96 and USE(``When did Gabriel Garcia die'', ``When did Gabriel Garcia \textit{Marquez} die'') = 0.91.\\

\textbf{Web search} results, while most accurately correlates with human judgment, also reflect sensitivity of the retrieval algorithm to the query formulation as well as the collection-specific selectivity of the query terms.
The resulting scores for our sample rewrites are R@10(``When is Robert Downey \textit{Jr} birthday'', ``When is Robert Downey \textit{Jrs} birthday'')=0.6 and R@10(``When did Gabriel Garcia die'', ``when did Gabriel Garcia \textit{Marquez} die'') = 0.78.\\

\section{Examples of Query Rewrites}
\label{appendix:rewrites}
In Table~\ref{tab:qr_rewrites} we show sample question rewrites from top 3 QR models along with conversational context.

\begin{table*}
\caption{A sample of conversation snippets from the QReCC test set with the question rewrites produced by Transformer++, CopyTransformer, and GECOR models.}
\centering
 \label{tab:qr_rewrites}
\scalebox{0.9}{
\begin{tabular}{l|l|l|l}
  \toprule
\textbf{Conversational context} & \textbf{Transformer++} & \textbf{CopyTransformer} & \textbf{GECOR} \\
 \toprule
\makecell[l]{\textit{Q1:} Did Nadia Comăneci \\ win any Gold medals in \\ the Olympics? \\ \textit{A1:} Nadia Comăneci is \\ a five-time Olympic gold \\ medalist. \\ \textit{Q2:} What about Silver?} & \makecell[l]{Did Nadia Comăneci \\ win any Silver medals?} & \makecell[l]{Did Nadia win \\ Silver medals?} & \makecell[l]{What about Silver \\ medals?} \\
\midrule
\makecell[l]{\textit{Q1:} What is range in \\ statistics? \\ \textit{A1:} The Range is the \\ difference between the \\ lowest and highest values. \\ \textit{Q2:} Describe some different \\ metrics to interpret it.} & \makecell[l]{Describe some different \\ metrics to interpret \\ range in statistics.} & \makecell[l]{Describe some different \\ metrics to interpret \\ range in statistics.} &  \makecell[l]{Describe some different \\ metrics to interpret \\ range.} \\
\midrule
\makecell[l]{\textit{Q1:} What is nominal GDP? \\ \textit{A1:} Nominal GDP is GDP \\ evaluated at current market \\ prices and includes all of the \\ changes in market prices \\ that have occurred during \\ the current year due to \\ inflation or deflation. \\ \textit{Q2:} What is the difference \\ with real?} & \makecell[l]{What is the difference \\ between nominal and \\ real GDP?} & \makecell[l]{What is the difference \\ with real GDP?} & \makecell[l]{What is the difference \\ with real GDP?} \\
\midrule
\makecell[l]{\textit{Q1:} Tell me about lavender \\ plants? \\ \textit{A1:} Lavandula is a genus of \\ 47 known species of flowering \\ plants in the mint family, \\ Lamiaceae. It is native to the \\ Old World and is found \\ from Cape Verde and \\ the Canary Islands, Europe \\ across to northern and eastern \\ Africa, the Mediterranean, \\ southwest Asia to southeast \\ India. \\ \textit{Q2:} What are the different \\ types?} & \makecell[l]{What are the different \\ types of lavender plants?} & \makecell[l]{What are the different \\ types of plants?} & \makecell[l]{What are the different \\ types of plants?} \\
\bottomrule
\end{tabular}}
\end{table*}

\section{Examples of Answers Found}
\label{appendix:answers_found}
In Table~\ref{tab:answer_examples} we provide two sample answers found by the baseline model.
In the first example, the baseline system picked the same passage as the human annotator, but extracted a different answer span from this passage.
In the second example, the baseline system picked a different passage than the human annotator.
While the produced answers are not exact matches to the corresponding human answers, we consider them as correct upon manual inspection.
This observation that a single question in QReCC can have multiple correct answers poses an important challenge for standard approaches to QA evaluation. 

\begin{table*}
\caption{A sample of answers produced by our end-to-end baseline for conversational QA. The baseline model can also produce relevant answers using spans that differ from the answers provided by the human annotators.}
\centering
 \label{tab:answer_examples}
\scalebox{0.9}{
\begin{tabular}{ll}
\toprule
\textbf{Human re-written question} & What are the educational requirements required to become a physician's assistant? \\[0.2cm]
\textbf{URL} & https://www.geteducated.com/careers/how-to-become-a-physician-assistant \\[0.2cm]
\textbf{Predicted URL} & https://www.geteducated.com/careers/how-to-become-a-physician-assistant \\[0.2cm]
\textbf{Human passage} & \ldots In most cases, a physician assistant will need a master's degree from an accredited \\
 & institution (two years of post-graduate education after completing a four-year degree). \\
 & \ldots Most applicants to PA education programs will not only have four \\
 & years of education, they will also have at least a year of medical experience. \\
 & \ldots five steps to becoming a PA: Complete your bachelor's degree (a science or \\
 & healthcare related major is usually best); Gain experience either working \\
 & or volunteering in a healthcare setting; Apply to ARC-PA accredited programs; \\
 & Complete a 2-3 year, master's level program; Pass the PANCE licensing exam. \\[0.2cm]
\textbf{Found passage} & (Same as human passage.) \\[0.2cm]
\textbf{Human answer} & Complete your bachelor's degree (a science or healthcare related major is usually best); \\
 & Gain experience either working or volunteering in a healthcare setting; \\
 & Apply to ARC-PA accredited physician assistant programs; \\
 & Complete a 2-3 year, master's level PA program; \\[0.2cm]
\textbf{Baseline model answer} & a physician assistant will need a master’s degree from an accredited institution \\
& (two years of post-graduate education after completing a four-year \\[0.2cm]
\textbf{Answer F1} & 15.38 \\[0.2cm]
\midrule
\textbf{Human re-written question} & What tools were used in the neolithic event? \\[0.2cm]
\textbf{URL} & https://sciencing.com/list-neolithic-stone-tools-8252604.html \\[0.2cm]
\textbf{Predicted URL} & https://stmuhistorymedia.org/neolithic-era-technology-advances-and-beginnings\\
& -of-agriculture \\[0.2cm]
\textbf{Human passage} & \ldots By the time the Neolithic came around, hand axes had fallen out of favor \\
& \ldots scientists consider the creation of all these tools a sign of early human ingenuity. \\
& Scrapers Scrapers are one of the original stone tools, found everywhere where people \\
& settled, \ldots Blades While a scraper can be used for cutting into an animal, a longer, \\
& thinner blade can be inserted deeper into a carcass, \ldots Arrows and Spearheads Arrows \\
& and spearheads are a more sophisticated shape than simple scrapers and blades. \ldots Axes \\
& The polished stone ax is considered one of the most important developments of the \\
& Neolithic era. \ldots Adzes The adze is a woodworking tool. \ldots Hammers and Chisels \\
& Chisels were made by attaching a sharp piece of stone to the end of a sturdy stick \ldots \\[0.2cm]
\textbf{Found passage} & \ldots The Neolithic Age was a period in the development of human technology, beginning \\
& about 10,000 BCE, in some parts of the Middle East, and later in other parts \\
& of the world, and ending between 4,500 and 2,000 BCE. \ldots Hunting also became much \\
& easier to accomplish with the introduction new of stone tools. The most common tools \\
& used were daggers and spear points, used for hunting, and hand axes, used for cutting up \\
& different meats, and scrappers, which were used to clean animal hides. \\[0.2cm]
\textbf{Human answer} & Scrapers. Scrapers are one of the original stone tools, found everywhere \\
& where people settled, long before the Neolithic Age began. ...Blades. ...Arrows \\
& and Spearheads. ...Axes. ...Adzes. ...Hammers and Chisels. \\[0.2cm]
\textbf{Baseline model answer} & The most common tools used were daggers and spear points, used for hunting, \\
& and hand axes \\[0.2cm]
\textbf{Answer F1} & 19.05 \\[0.2cm]
\bottomrule
\end{tabular}}
\end{table*}

\end{document}